# Speed and Accuracy of Static Image Discrimination by Rats

Pamela Reinagel[1,*] and Robert E Clark[2,3]

1. Section of Neurobiology, Division of Biological Sciences
2. Veterans Affairs Medical Center, San Diego, CA 92161, USA
3. Department of Psychiatry

University of California, San Diego, La Jolla, CA 92093, USA

*Correspondence: preinagel@ucsd.edu

## Abstract

**When discriminating dynamic noisy sensory signals, human and primate subjects achieve higher accuracy when they take more time to decide, an effect attributed to accumulation of evidence over time to overcome neural noise. We measured the speed and accuracy of twelve freely behaving rats discriminating static, high contrast photographs of real-world objects for water reward in a self-paced task. Response latency was longer in correct trials compared to error trials. Discrimination accuracy increased with response latency over the range of 500-1200ms. We used morphs between previously learned images to vary the image similarity parametrically, and thereby modulate task difficulty from ceiling to chance. Over this range we find that rats take more time before responding in trials with more similar stimuli. We conclude that rats' perceptual decisions improve with time even in the absence of temporal information in the stimulus, and that rats modulate speed in response to discrimination difficulty to balance speed and accuracy.**

## Introduction

The temporal dynamics of decision making are often studied in the context of temporally dynamic stimuli. For example in the random dot motion task, a number of randomly positioned dots move coherently in one of two directions, while a number of other randomly positioned dots move in random directions. Thus the signal (direction of coherent motion) is embedded in noise, and averaging over time improves the signal-to-noise ratio of the sensory information. Here we focus on two basic observations from that paradigm: (1) when making difficult discriminations in this task, accuracy improves with viewing time; (2) subjects wait longer to respond when the stimuli are less coherent and therefore more ambiguous (Palmer et al., 2005). An extensive experimental and theoretical literature relates the timing and accuracy decisions of behaving humans and non-human primates to parameters of the stimulus and reward, and the neurophysiology of neurons involved in sensory, motor, and choice aspects of the task (Bogacz et al., 2009; Gold and Shadlen, 2007). A key feature of most models is that the improvements in decision accuracy with time reflect the integration of sensory signals over time to overcome noise.  In the random dot motion task, it is thought that the relevant noise sources are internal to the brain, and not related to the dynamic or stochastic nature of the stimulus (Britten et al., 1993). If this is the case, similar results should be found for sensory discriminations between static, high-contrast visual images, in which the difficulty of the discrimination is determined by the similarity between the images rather than by any noise component.

Relatively little is known about the trade off of speed and accuracy in perceptual decisions by rats or rodents in general. In the olfactory modality, discrimination tasks can be made arbitrarily difficult by making different mixtures of two discriminable odors. In a task where rats sample odor stimuli over time by sniffing, it was found that rats' performance does not improve with time, arguing against dynamic models of information processing or accumulation of sensory evidence with time in the olfactory system (Kepecs et al., 2006; Uchida et al., 2006; Uchida and Mainen, 2003). In a different paradigm, the accuracy with which mice discriminate highly similar odors improves with exposure time, providing evidence in favor of speed-accuracy trade-off and temporal integration of information (Abraham et al., 2004; Rinberg et al., 2006). The role of temporal processing in olfaction is likely to be quite different from that of the visual system, however. In the random dot visual motion task, there is evidence that rats' accuracy improves with viewing time (Reinagel et al., 2009; Reinagel et al., 2012). Here we extend this literature by reporting the speed and accuracy of sensory decision making by rats discriminating between static visual images with varying degrees of visual similarity.

## Materials and Methods
*Behavioral training and testing*

Twelve female Long-Evans rats (Harlan) were water restricted and trained to perform visual tasks for water reward (Meier, Flister & Reinagel 2011). Subjects began training at age p30 for 2hrs/day 7 days a week. Subjects performed 500-1500 trials per day, and received water in 50% of trials when performing at chance. Reward magnitude was adjusted moderately for each subject to ensure adequate total water consumption, but no supplemental water (outside of the task) was given at any time. Hydrating treats (carrots) were given after each training session, however. During training sessions subjects had free access to return to the home cage at any time; thus they had access to food during periods of water consumption. On this protocol, all subjects maintained normal growth curves (within 5% of published values for unrestricted food and water). Between training sessions, subjects were pair-housed with enrichment (chew toys, PVC tubes). Subjects were housed in a reverse 12 hour light/dark cycle and were trained and tested in the housing environment during the dark cycle. All procedures were performed with the approval and under the supervision of the UCSD IACUC, within an ALAAC accredited animal facility.

In preliminary shaping, subjects moved through four shaping steps (Table 1) to acquire a two alternative forced choice (2AFC) visual discrimination between static grayscale photographic images of two real world objects (a statue and a space shuttle). In this and all subsequent steps, each trial was initiated by the subject by licking a central request port, which caused the two images to appear on the screen, one above each response port. The rewarded (S+) stimulus was randomly assigned to either the left (L) or right (R) side of the screen, and the unrewarded (S-) stimulus to the other side. The two images were large and high contrast, and were matched in luminance, size, contrast, and orientation. The images persisted until the subject licked a response port (L or R), with no time limit. Responses at the port co-localized with the S+ stimulus were rewarded with water delivered at the same location with <10ms delay, after which the subject could immediately initiate a new trial. Responses at the port co-localized with the S- stimulus were penalized with a timeout of 2-10sec before a new trial could be initiated.

After each correct trial, the S+ stimulus was assigned to L or R side with equal probability. After an error trial, however, there was a fixed probability (0.25-0.5) of entering a correction trial instead, in which case the S+ stimulus was deterministically placed at the port opposite the previous trial's response. This method was highly successful in helping rats overcome bias (overall preference for one response port over the other) as well as perseveration (preference to return to the most recently visited or recently rewarded port) over months of automated training and testing. However it alters the statistics of the task in trials after errors. Therefore only trials after correct trials are analyzed here.

Reward magnitude (water volume) was empirically adjusted for each rat to ensure adequate hydration and normal growth curve, while maximizing motivation as judged by the number of trials per day. Penalty time out duration was empirically adjusted for each rat to discourage guessing, while avoiding excessive subject frustration as judged by quitting. Both reinforcement parameters remained fixed for each rat within each training session, and were adjusted infrequently over the rat's lifespan.

After mastering the first 2AFC visual discrimination (Shaping Step 4), subjects learned a second visual discrimination between two novel images (a paintbrush and a flashlight), one of which was assigned to be the S+ stimulus for each rat (Shaping Step 5). Subjects were trained on this "exemplar" discrimination until performance exceeded 80% accuracy for at least 200 trials before entering the test phase (Shaping Step 6).

In the test phase, subjects continued to be tested on the exemplar discrimination in 80% of trials; later analysis confirmed that performance on the exemplar pair was asymptotic and stationary for the duration of the test phase. In the remaining 20% of trials (interleaved), subjects were presented with a pair of images of parametrically varied similarity, obtained by morphing between the S+ and S- exemplar images. In these probe trials, subjects were rewarded for responding at the port co-localized with the stimulus that was closer to S+ of the two images. The order of probe trial types was pseudorandom with the constraint that each of the 14 non-exemplar difficulty levels had to be presented once before any one difficulty level could be repeated. This procedure ensured that data for probe trials accrued at the same rate for every difficulty level. Each rat continued the test phase until each probe type was tested exactly 150 times.

All 12 rats that began the study learned the task and completed the study. The total training time in calendar days from naive animal to beginning the testing phase (Shaping Steps 1-5) ranged from 29-108 days (56.1+/-26.3, mean+/-std), corresponding to ages between p59-p138. The calendar days required to complete the testing period (Step 6) ranged from 20- 42 days (27.4+/-5.9, mean+/-std). The shaping steps are summarized in Table 1; additional details on the training paradigm are described in (Meier, Flister and Reinagel 2011).

| Shaping step | Description | Days to complete (min-max) |
|---|---|---|
| 1. Free Drinks | Water released at any port when triggered by licking, and also un-triggered at random times | 0-4 |
| 2. Earned Drinks | Water at any port when triggered by licking only; requires rotating among all three ports | 0-9 |
| 3. Approach Visual Target, 2AFC | Upon request (licking unrewarded center port), S+ (statue) image appears over one response port; responses at S+ rewarded with water, response on other side (no image) penalized with timeout | 4-11 |
| 4. Visual Discrimination, 2AFC | Upon request S+ (statue) image appears over one response port and matched S- (space shuttle) over the other. Responses at S+ rewarded with water, response at S- penalized with timeout | 16-43 |
| 5. Exemplar Discrimination, 2AFC | Same as previous, but S+ is now either flashlight or paintbrush, and S- is the other image of this pair | 29-108 |
| 6. Testing: Exemplar and Probes, 2AFC | Same as previous, but 20% of trials are probes with morphed intermediates between S+ and S- | 50-141 |

**Table 1. Behavioral shaping strategy for parametric image discrimination task.**

*Data Analysis*
Each behavioral training or testing block is defined by the statistics of the stimulus ensemble (constant over a training step); the magnitude of reward on correct trials (constant within block); and the duration of the time-out on error trials (constant within block). Each trial is further defined by the specific visual stimuli (selected independently each trial); and the side on which the target appeared and response was rewarded (selected independently each trial, except for correction trials, which are therefore excluded from analysis). In each trial we recorded these variables as well as the time of subject-initiated stimulus request, the latency from stimulus onset to response, and the outcome of the trial (correct/reward or error/timeout). Data analysis was performed using Matlab (Mathworks, Natick MA).

# Results

We trained twelve Long-Evans rats to discriminate between static photographs of two perceptually similar objects – a flashlight and a paintbrush – in a self-paced 2AFC operant conditioning paradigm (Figure 1a-d; for details see Methods and Table 1). After performance was asymptotic on this "exemplar" discrimination, subjects began the testing phase. During testing, the exemplar discrimination was tested in 80% of trials; the remaining 20% of trials were probe trials in which the discriminated images were rendered more similar by morphing between the exemplar images (Figure 1c). In probe trials, subjects were rewarded for selecting the image that more closely resembles the learned target.

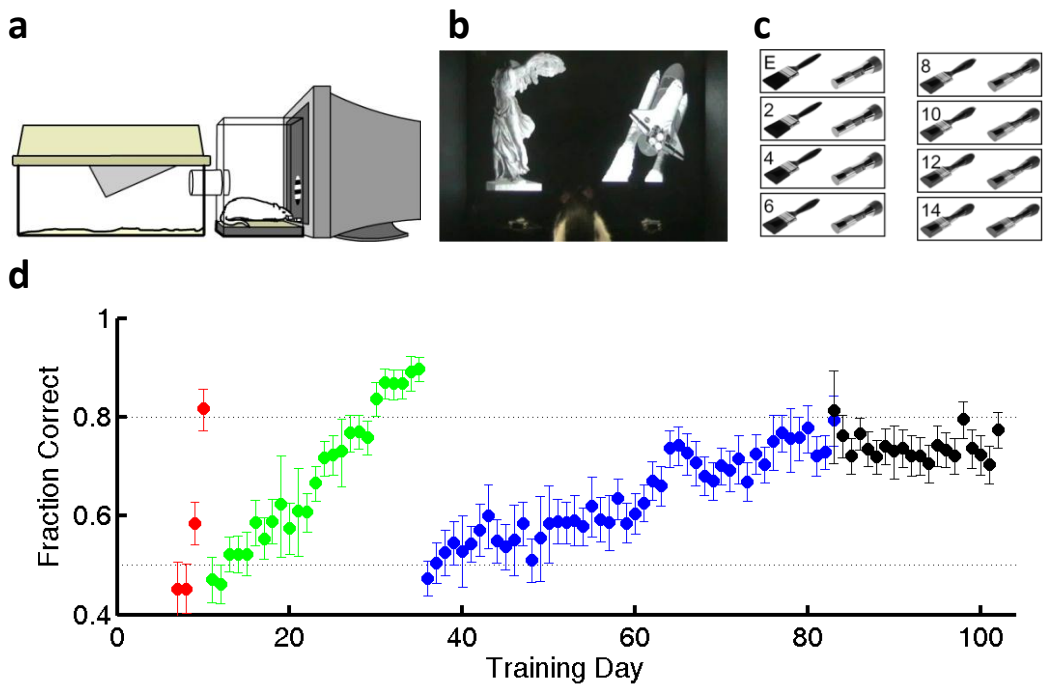

**Figure 1. Training and testing paradigm. a)** Diagram of cage-attached operant conditioning chamber. **b)** One of the subjects in this study in the operant chamber performing the statue-shuttle image discrimination (Shaping Step 4). **c)** The exemplar image pair E and examples of the intermediate morph pairs for the flashlight-paintbrush image discrimination used in the testing phase (Shaping Step 6). **d)** Example learning curve for one subject showing performance as a course of training from naïve to study completion. Training day indicates number of calendar days since initiating training. Chance performance is 0.5 (lower dotted line). In the first two shaping steps (acclimation to apparatus, Shaping Steps 1-2, days 1-6 in this case) all responses are valid, so performance is undefined (not plotted). For all subsequent shaping steps, each symbol shows the average performance on one task over one training day. Error bars show 95% binomial confidence intervals. Color indicates task: go to statue (Shaping Step 3, *red*), discriminate statue from shuttle (Shaping Step 4, *green*), discriminate flashlight from paintbrush exemplars (Shaping Step 5, *blue*), or discriminate flashlight from paintbrush including exemplars and morph probe trials (Shaping Step 6, *black*). Subjects were automatically graduated to the next task when performance exceeded 80% (upper dotted line) for at least 200 trials, and graduated from the final task when each morph level had been tested exactly 150 times.

## Response Latency Longer in Correct Trials

For each trial we define the latency of the response as the time between voluntary initiation of the trial (at which time images appear) and the subject's response (at which time the images disappear and reward or penalty occurs). The distribution of response latency for exemplar discriminations is shown for one rat (Figure 2a and 2b) for both correct trials and for error trials. For this subject, short latencies (500-1000ms) are more frequent in error trials than correct trials, while long latencies (1000-1500ms) are more frequent in correct trials than error trials. The median response latency was longer in correct trials than error trials for this subject (arrows in Figure 2a,b), and for all 12 subjects (not shown; $P<10^{-3}$ by Wilcoxon signed rank test).

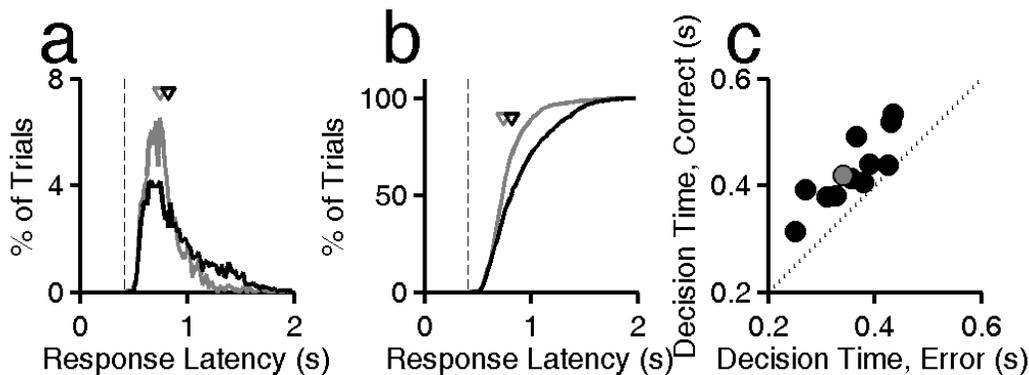

Figure 2 Longer latencies in correct trials. **a)** Distribution of response latency for error trials (black) and correct trials (gray) for exemplar discriminations for one subject (same subject as Figure 1d). Arrows indicate median latencies of the distributions. Dashed line is the minimum response latency this subject showed in any trial or task. **b)** Cumulative distributions of response latency, the integrals of curves shown in a. **c)** Median decision time in error trials (x axis) and in correct trials (y axis) for each subject (N=12), for exemplar discriminations in the test phase. Subjects trained and tested in station 1 (circle) or station 2 (square) showed similar results. The example subject used in a and b is highlighted (gray). Symbols are above the identity line (dotted) if correct trials had longer median response latency.

The minimum response latency of a given subject across all trials and all visual 2AFC tasks (dashed line, Figures 2a and 2b) places an upper bound on the time required for retinal processing and motor response for that subject. This delay was stable over time for a given subject and ranged from 323ms to 413ms across subjects. We define the "decision time" in each trial as the response latency minus the subject's sensory/motor delay as defined above. The median decision time for correct trials was longer than in error trials for all 12 subjects (Figure 2c; $P<10^{-3}$ by Wilcoxon signed rank test).

## Dependence of Accuracy on Response Latency

The fact that response latencies tend to be longer in correct trials implies that accuracy (% correct) was higher in trials with longer response latency. The relationship between response latency and accuracy on exemplar trials is shown for an example subject in Figure 3a, and summarized for all subjects in Figure 3b. Performance improved with response latency over the range of 500ms to 1200ms, beyond which further response delay did not help, despite the fact that performance remained below 100%. If we define trials with response latency in that subject's lowest quartile as "fast" and those with response latency in the highest quartile as "slow", rats performed better in slow trials than fast ones (Figure 3c). All subjects showed an advantage of waiting before responding; this was significant for 10/12 rats individually (the 95% binomial confidence intervals do not overlap), and the effect was significant at the population level ($P<10^{-3}$, Wilcoxon signed rank test).

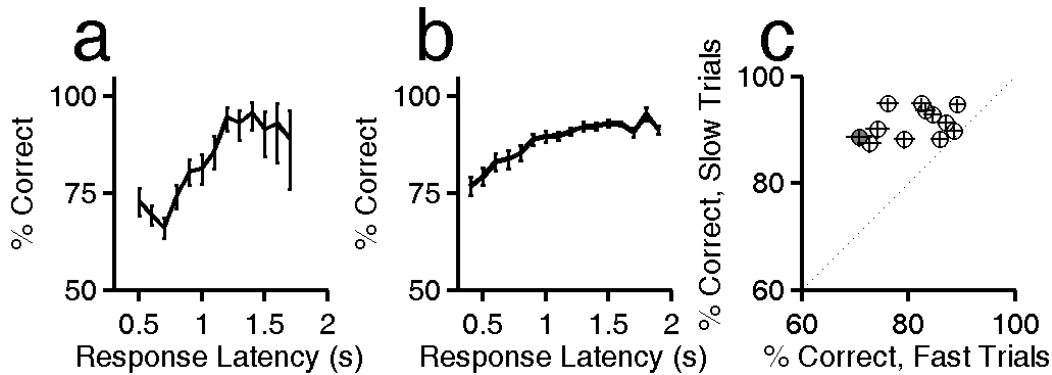

**Figure 3 Accuracy improves with response latency. a)** Accuracy of exemplar discrimination as a function of response latency for a single subject (same rat as Figure 1d and 2ab); error bars show the 95% binomial confidence intervals. **b)** Accuracy of exemplar discrimination as a function of response latency averaged over all N=12 rats; error bars show SEM over the population. **c)** Accuracy on exemplar discrimination in fast trials vs. in slow trials in the test phase. Each symbol represents data from a single rat, and error bars show 95% binomial confidence intervals. Subjects trained and tested in station 1 (circle) or station 2 (square) showed similar results. The example subject used in a and b is highlighted (gray). Symbols are above the identity line (dotted) if slow trials had higher accuracy.

### Rats take more time to decide when discrimination is more difficult

The image discrimination task is self-paced: trials occur only when initiated by the subject. There is no time limit to respond, but most trials are completed within 2s of stimulus onset. To test whether rats take longer to make a decision when the sensory stimuli are more ambiguous, we parametrically varied the similarity of the two images in probe trials with morphed images (Methods; Figure 1c)(Clark et al.).

Accuracy of discrimination decreased as the images became more similar, as shown for one rat in Figure 4a and summarized for all rats in Figure 4b. For the subject whose performance is shown in Figure 4a, the distribution of response latencies was shifted to longer latencies in the trials with more ambiguous stimuli (Figure 4c), consistent with the hypothesis that this subject took more time on more difficult trials. For most subjects (N=10/12 rats), the median latency on the easiest trials (exemplar, level 1) was lower than the median latency on the most difficult or ambiguous trials (morph levels 12-15) (Figure 4d), and this trend was significant at the population level ($P<10^{-2}$, Wilcoxon signed rank test).

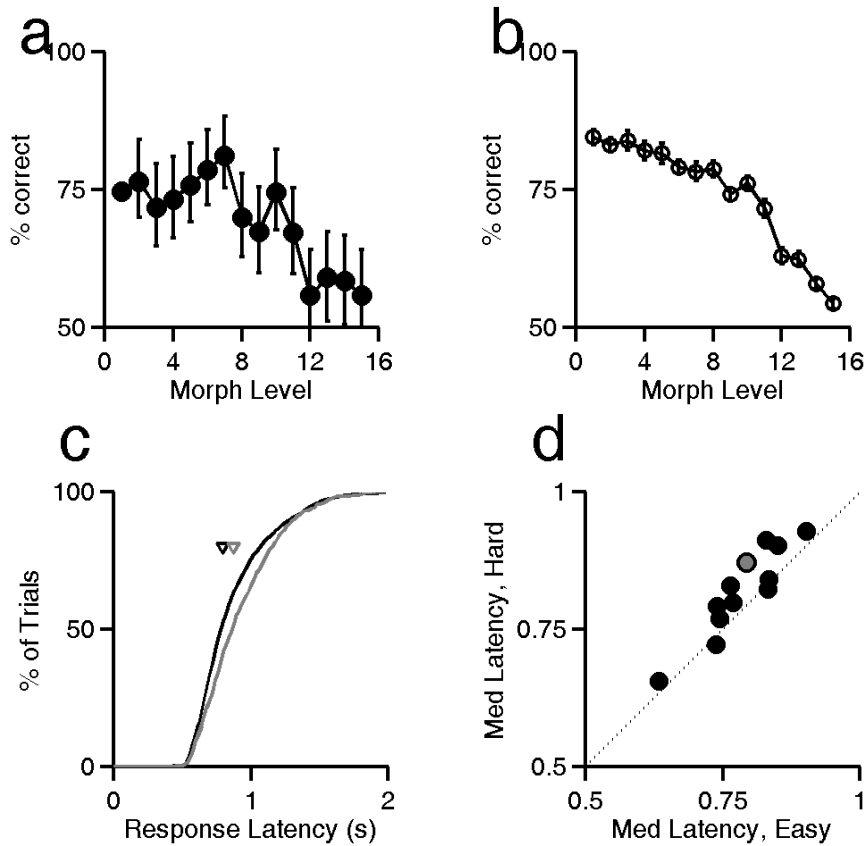

**Figure 4. Response latency increases with trial difficulty. a)** Performance (% correct responses) as a function of stimulus ambiguity (morph level) for one rat (cf. Figs. 1d, 2a-b, and 3a). Error bars show 95% binomial confidence intervals. **b)** Average performance of all 12 subjects as a function of stimulus ambiguity. Error bars show SEM over the population of N=12 subjects. **c)** Cumulative distribution of response latency for the subject analyzed in a, for the easiest (level 1, black curve) and hardest (levels 12-15, gray curve) trials. Arrows indicate the median latencies of the two distributions. **d)** Median latency for easiest and hardest trial types for all N=12 rats; data for the subject shown in panels a and c is highlighted in gray. Symbols above the diagonal unity line indicate a subject that takes more time to decide on harder discriminations.

## Discussion

Our results show that rats exhibit speed-accuracy trade-off in visual discriminations: the longer rats waited to decide, the more accurate their decisions were (Figure 3). This establishes a correlation between accuracy and response latency, but does not establish causation. It may be that rats achieved better accuracy because they took more time, but we cannot exclude alternative causal models. For example, a hidden state like attention or motivation could affect both response latency and accuracy. Future experiments could differentiate among these possibilities by imposing mandatory waiting times or response deadlines, or manipulating neural circuits underlying executive control.

Most rats also waited longer to make decisions between more similar stimuli (Figure 4), as predicted if they were integrating evidence to a decision threshold. The additional time was not sufficient to achieve equal performance on all difficulty levels, as would be predicted by a strict decision threshold model. This may be because there is a limit to how long rats can integrate information, beyond which more time does not help (Figure 3a,b). Future experiments could test this hypothesis by measuring the improvement in accuracy as a function of stimulus duration (under experimental control), measuring the stability of information as a function of waiting time after stimulus extinction; and by measuring the effect of speed on accuracy as a function of stimulus difficulty.

The "failure" of rats to wait longer may also reflect a strategy for reward harvesting, in light of the opportunity cost of waiting. Incorrect responses result in timeout, delaying the next possible reward substantially. But even a completely uncertain subject has a 50% probability of being correct by chance, in which case she will receive a reward immediately, and proceed immediately to a new trial. Moreover, if the current discrimination is difficult, it is likely that the next one will be easier (because probe trials were rare). It is only worth waiting longer if it will increase the chance of a correct answer sufficiently to compensate for the expected delay in the reward, as well as the expected delay in beginning the next trial. Future experiments could test this hypothesis by manipulating the opportunity cost, for example changing the time-out delay, enforcing a fixed inter-trial interval, and measuring the subjective temporal discounting of the reward for each rat. Nevertheless our data are consistent with the framework of integration to a decision threshold, if this threshold decays over time.

Our results provide evidence for a time-dependent process of sensory decision making in image discrimination, despite the absence of dynamics or noise in the stimulus (Figure 3). What is this time-dependent process?

Humans and other primates scan static visual images with saccades and fixations to extract information over time (Yarbus, 1961). We did not track head and eye movements in this task, but we think active sampling in the sense of scanning the image is unlikely to play a role in our task. Rats do not have fovea, and typically only move their eyes once every several seconds (Fuller, 1985; Hikosaka and Sakamoto, 1987; Meier, 2011). With response latency typically less than one second, rats probably don't make more than one fixation during a trial.

Even during a single fixation, there are small eye movements which convert spatial patterns in static stimuli into a spatiotemporal patterns on the retina, with important consequences for visual information extraction (Engbert et al., 2011; Hennig and Worgotter, 2007; Kagan et al., 2008; Kuang et al., 2012; Martinez-Conde et al., 2004; Martinez-Conde et al., 2009; Rucci, 2008; Rucci et al., 2007). This effect helps explain why stabilized retinal images fade perceptually. Nevertheless, these important temporal fluctuations in the periphery may not result in significant temporal fluctuations in the responses of higher level visual neurons with much larger receptive fields, such as might be responsible for discriminating a flashlight from a paintbrush.

Our data are consistent with the hypothesis that accuracy improves with time as the sensory signal is integrated to overcome intrinsic noise in the firing of these higher level neurons. For example, suppose the firing rate of an object-selective visual neuron encodes the match of the stimulus in its receptive field to its target. Even if the drive (input) to the neuron is constant, its rate must be inferred by downstream neurons from a stochastic spike train. Longer integration times will yield more accurate estimates -- especially when the signal is weak (match is poor) and thus firing rate is low.

In primates performing the random dot motion task, the direction of motion is encoded in cortical area MT (Britten et al., 1996; Britten et al., 1992, 1993). The integration of evidence for competing hypotheses (L or R motion) is thought to occur in the downstream area LIP (Gold and Shadlen, 2001; Shadlen and Newsome, 1996, 2001), among other places, depending on motor response mode. We do not know where in the brain of the rat the differences between complex images are encoded. The match of each displayed image to the target (S+) or distractor (S-) image could plausibly be encoded in the rat homolog of primate area TE. There is behavioral evidence that rats possess transformation-invariant object recognition (Tafazoli et al.; Zoccolan et al., 2009), but the neurons responsible for encoding this representation have yet to be identified. The perirhinal cortex lies downstream of TE, and has been implicated in subtle sensory discriminations between similar objects (Buckley et al., 2001; Buckley and Gaffan, 1998; Bussey et al., 2002, 2003), though we note that perirhinal cortex is not required for this task (Clark et al.). The site of the putative decision neurons is unknown. Future experiments could constrain candidates by using a more restricted motor response.

In summary, our data reveal a speed-accuracy trade-off in the visual discrimination of images of natural objects by rats. Rats performed better when they chose to take more time to decide (Figures 2 and 3), despite the absence of any temporal information in the stimulus itself. Moreover, most rats took more time to make a decision when confronted with more difficult discriminations (Figure 4), consistent with the hypothesis that rats can modulate response speed in the service of increasing accuracy.